\documentclass[%
aip,
amsmath,amssymb,
reprint,%
]{revtex4-1}

\usepackage{graphicx}
\usepackage{dcolumn}
\usepackage{bm}

\usepackage[utf8]{inputenc}
\usepackage[T1]{fontenc}
\usepackage{mathptmx}

\usepackage{graphicx}
\usepackage{dcolumn}
\usepackage{bm}
\usepackage{mathrsfs}
\usepackage{natbib}
\usepackage{amsmath,esint}
\usepackage{epstopdf}
\usepackage{color}
\usepackage{placeins}
\usepackage{amsmath}
\usepackage{mathtools}
\usepackage{float}
\usepackage{refcount}
\usepackage[english]{babel}

\usepackage[mathlines]{lineno}

\newsavebox{\astrutbox}
\sbox{\astrutbox}{\rule[-5pt]{0pt}{20pt}}

\newcommand\rvec{\textbf{\textit{r}}}

\newcommand\xvec{\textbf{\textit{x}}}

\newcommand\uvec{\textbf{\textit{u}}}
\newcommand\fvec{\textbf{\textit{f}}}

\def\be{\begin{eqnarray}}
	\def\ee{\end{eqnarray}}
\def\bes{\begin{subeqnarray}}
	\def\ees{\end{subeqnarray}}

\def\lp{\left(}
\def\rp{\right)}
\def\lb{\left[}
\def\rb{\right]}
\def\lcb{\left\{}
\def\rcb{\right\}}
\def\befi{\begin{figure}}
	\def\eefi{\end{figure}}

\def\bce{\begin{center}}
	\def\ece{\end{center}}

\def\lap{\nabla^2}

\def\ba#1\ea{\begin{align}#1\end{align}}

\def\bsa#1\esa{\begin{subequations}
		\begin{align}#1\end{align} \end{subequations}}

\usepackage[dvipsnames]{xcolor}
\definecolor{mycolor}{rgb}{0.0,0.2,0.6}

\def\hl{\textcolor{black}}

\begin{document}

\preprint{AIP/123-QED}

\title[Stealthy Movements and Concealed Swarms of Swimming micro-Robots]{Stealthy Movements and Concealed Swarms of Swimming micro-Robots}

\author{Mehdi Mirzakhanloo}
\affiliation{Department of Mechanical Engineering, University of California Berkeley, CA 94720, USA } %

\author{Mohammad-Reza Alam}
\email{reza.alam@berkeley.edu}
\affiliation{Department of Mechanical Engineering, University of California Berkeley, CA 94720, USA } %

\date{\today}

\begin{abstract}
	
Here we show that micro-swimmers can form a \textit{concealed} swarm through synergistic cooperation in suppressing one another's disturbing flows. We then demonstrate how such a concealed swarm can actively gather around a favorite spot, point toward a target, or track a desired trajectory in space, while minimally disturbing the ambient fluid. Our findings provide a clear road map to control and lead flocks of swimming micro-robots in \textit{stealth} versus \textit{fast} modes, tuned through their active collaboration in minimally disturbing the host medium.

\end{abstract}

\maketitle

\hl{\section{Introduction}}

Swimming micro-robots capable of navigating through fluid environments are at the forefront of minimally invasive therapeutics and theranostics \cite{nelson2010microrobots}. They hold great promise for a wide range of biomedical applications including targeted drug delivery, micro-surgery, remote sensing and localized diagnostics \cite{nelson2010microrobots,peyer2013bio,li2017micro,erkoc2019mobile}. The past decade has seen a great leap forward in science and engineering of these miniaturized untethered robots \cite{sitti2017mobile}. Particularly, remarkable progress has been made toward exploring various propulsion mechanisms \cite{abbott2009should,bente2018biohybrid,alapan2019microrobotics}, design and fabrication approaches \cite{peters2016degradable}, imaging technologies for real-time motion tracking \cite{pane2019imaging}, and manipulation techniques for navigation and motion control \cite{martel2009flagellated,kummer2010octomag,khalil2013magnetic,xie2016controlled}.

However, \textit{optimal} strategies in swarm control remain largely unexplored for swimming micro-robots \cite{bente2018biohybrid}. As a result, little is understood about their potential ability as a \textit{group} to optimize their fitness and functionality. A few recent studies have only shown a glimpse of such potentials in the realm of micro-scale swimmers. For instance, actively controlled cooperation between artificial micro-swimmers has been reported to significantly improve (both the capacity and precision of) micro-manipulation and cargo transport \cite{huang2014cooperative}. It has also been recently shown \cite{mirzakhanloo2018hydrodynamic} that a pair of interacting micro-swimmers can boost each other's swimming speed through \hl{ambient fluid}. This observation (termed as `hydrodynamic slingshot effect') implies that by forming a swarm, swimming micro-robots can collaborate and travel faster as a group than single individuals. Now, the more intriguing question is whether by forming a swarm, swimmers are also able to smartly cancel out each other's disturbing effects to the fluid environment. In other words, is it possible to form a stealth swarm minimally disturbing the ambient fluid? And if so, to what extent such cooperation between the agents can be effective in stifling the swarm's hydrodynamic signature?

Here we unveil synergistic cooperation of micro-swimmers (in suppressing one another's disturbing flows) that leads to the formation of stealth swarms. We refer to this mode of swarming as the \textit{concealed} mode, which can reduce the swarm's net induced disturbances by more than 99\% (or 50\%) in three-dimensional (or two-dimensional) movements. This is equivalent to quenching the swarm's hydrodynamic signature (and thus shrinking its associated detection region) by an order of magnitude in range. Through numerical experiments, we then demonstrate how such a concealed swarm can actively gather around a favorite spot, point toward a target, or track a desired trajectory in space, whilst minimally disturbing the surrounding environment.


\hl{\section{ Problem Formulation \& Approach}} \label{sec.theory}

\hl{Dynamics of the incompressible flow around swimming objects is governed by the Navier-Stokes equations: 
\ba\label{eqA1}
\rho \frac{D \uvec}{D t} =  -\nabla P + \eta \ \lap \uvec + \bm{F}, \quad \nabla \cdot \uvec = 0,
\ea
subject to boundary conditions imposed by their body deformations. Here, $\rho$ and $\eta$ are density and dynamic viscosity of the surrounding fluid, $P$ denotes the pressure field, $\uvec$ is the velocity field, and $\bm{F}$ represents the external body force per unit volume. The relative importance of inertial to viscous effects can also be quantified by the Reynolds number, Re$=\rho$UL/$\eta$, where U and L denote characteristic velocity and length, respectively. For micro-scale swimmers (also known as \textit{micro-swimmers}) swimming in water ($\rho \approx 10^3$ kg$/$m$^3$ and $\eta \approx 10^{-3}$ Pa.s) the corresponding Reynolds number is always very small (i.e., Re $\ll 1$). Common examples include: (i) typical bacteria, such as \textit{Escherichia coli}, with length of $\sim$ 1-10 $\mu$m and swimming speed of $\sim$ 10 $\mu$ms$^{-1}$ \cite{darnton2007torque}, for which the Reynolds number is $\sim$ $10^{-5}$-$10^{-4}$ when swimming in water; or (ii) the green algae \textit{Chlamydomonas reinhardtii} with characteristic length L $\sim$ 10 $\mu$m and swimming speed U $\sim$ 100 $\mu$ms$^{-1}$ \cite{goldstein2015green}, which result in the Reynolds number Re $\sim$ $10^{-3}$. Thereby, it is appropriate to study micro-swimmers in the context of low Reynolds number regimes (Re $\ll 1$), where the fluid inertia is negligibly small compared to the fluid viscosity, and the viscous diffusion dominates fluid transport. The Navier-Stokes equations then simplify to the Stokes equation:}
\ba\label{E1}
\nabla P = \eta \ \lap \uvec + \bm{F}, \quad \nabla \cdot \uvec = 0,
\ea
\hl{which has no explicit time-dependency.} This along with its linearity, makes the Stokes equation invariant under time-reversal. As a result, sequence of body deformations (or swimming strokes) that are reciprocal (i.e. invariant under time-reversal), do not generate a net motion at Stokes regime. This means that typical swimming strategies used by larger organisms (e.g. fish, birds, or insects), are not effective at micro- and nano-meter scales. Therefore, motile microorganisms have evolved alternative propulsion mechanisms to break the time-symmetry, while retaining periodicity in time \cite{purcell1977life}. Their swimming strategies are often based on drag anisotropy on a slender body in Stokes regime, and include cork-screw propulsion of bacteria, flexible-oar mechanism of spermatozoa, and asymmetric beats of bi-flagellate alga \cite{purcell1977life}.

These inherently available natural micro-swimmers, not only inspire the design of fully synthetic micro-robots \cite{bente2018biohybrid}, but also may be functionalized and directly used as steerable swimming micro-robots \cite{alapan2019microrobotics}. Here we are interested in \textit{flocks} of such bio-inspired and bio-hybrid swimming micro-robots, where each agent is either a real microorganism or meticulously synthesized to mimic one. Therefore, to model their induced disturbances, we will treat each individual swimmer as a swimming microorganism.

\begin{figure}[!htb]
	\centering 
	{\includegraphics[width=0.47\textwidth]{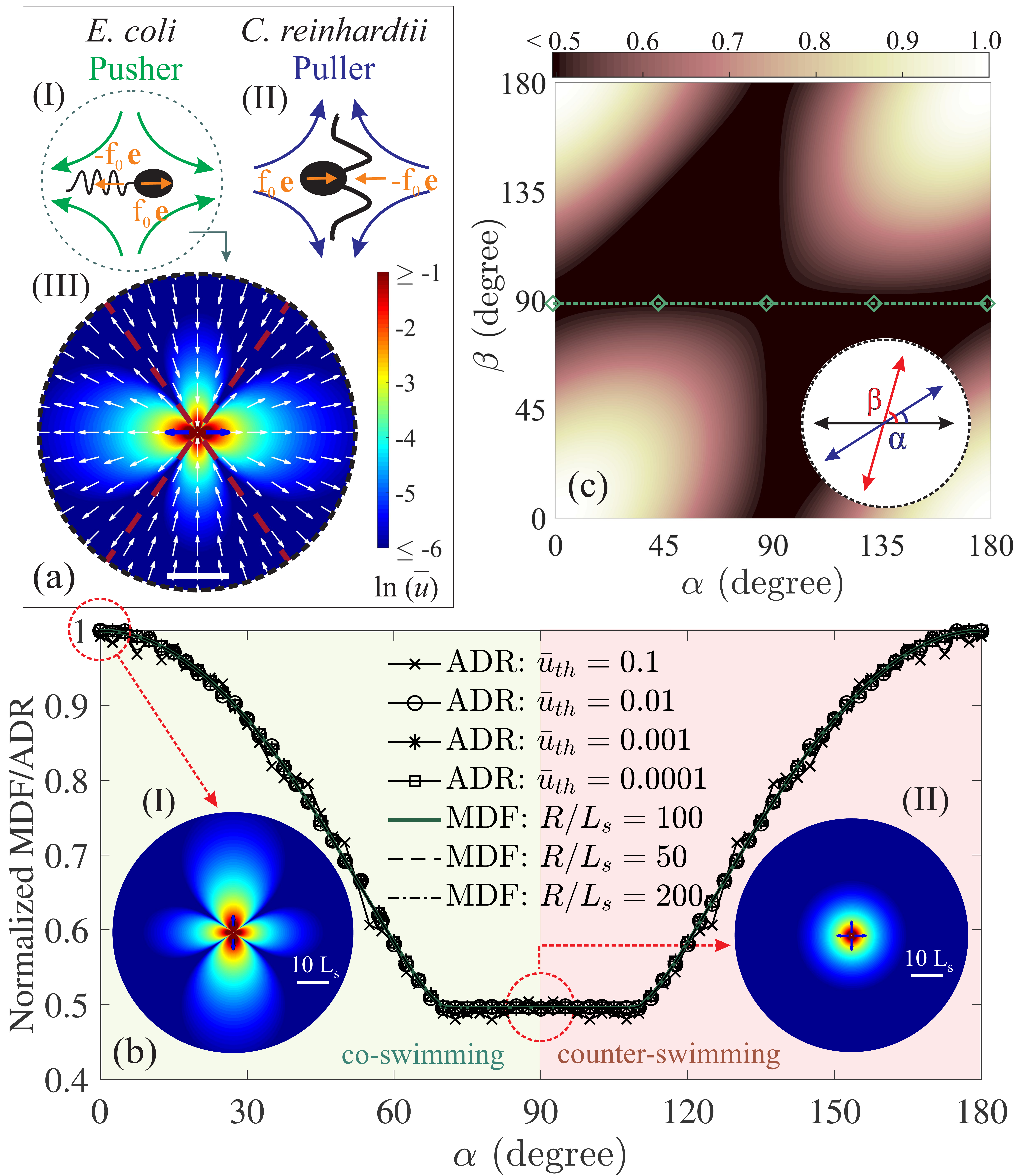}}
	\caption {
		\textbf{(a):} Schematic representation of archetypal pusher (I) and puller (II) micro-swimmers. Forces exerted by each swimming microorganism to the ambient fluid are represented, and induced flow directions are shown by curly arrows. The flow field induced by an extensile force dipole (as a model for pushers) is presented as the inset (III), where color shading represent the magnitude and \hl{white} arrows show direction of the induced flow field. \hl{The scale bar denotes $10L_s$.}
		\textbf{(b):} Fluid disturbances induced by a group of two swimmers are measured in terms of MDF/ADR, and plotted as a function of relative angle between the agents ($\alpha$). Values presented in each plot are normalized by their maximum, which corresponds to the reference case of aligned swimmers (i.e. schooling arrangement). We also compare normalized values of MDF computed over surrounding rings of various radii, specifically $R/L_s$ = 50, 100, and 200. Similarly, the process of computing ADR has been repeated for different threshold values. Specifically we present those corresponding to $\bar{u}_{th}$ = 0.1, 0.01, 0.001, and 0.0001. Insets: Magnitude of the flow induced by two micro-swimmers swimming in parallel (I) or perpendicular (II) directions.
		\textbf{(c):} Fluid disturbances (in terms of MDF) induced by a group of three swimmers is presented by color shading over the space of relative angles ($\alpha$, $\beta$) between them. The values are computed over a surrounding ring of radius $R/L_s = 100$, and then normalized by the reference case (i.e. their maximum value). Inset schematically defines $\alpha$ and $\beta$.}
	\label{figM1}
\end{figure}

\hl{In Stokes regime, self-propelled buoyant micro-swimmers exert no net force and no net torque to the ambient fluid. Flagellated microorganisms, for instance, use their flagella --flexible  external appendages-- to generate a net thrust, and propel themselves through ambient fluids. This propulsive force --generated mainly owing to the drag anisotropy of slender filaments in Stokes regime \cite{batchelor1970slender}-- is, however, balanced by the drag force acting on the cell body (see e.g. Fig. \ref{figM1}a). Hence,} in the most general form, far-field of the flow induced by each micro-swimmer can be well described by the flow of a force dipole. \hl{To be more precise, by the flow of a force dipole composed of the thrust force generated by swimmer's propulsion mechanism, and the viscous drag acting on its body. Note that the model dipole is contractile for swimmers with front-mounted flagella (i.e., `pullers' such as \textit{C. reinhardtii}), and extensile for those with rear-mounted flagella (i.e., `pushers' such as \textit{E. coli}). Schematic representations of the force dipoles generated by archetypal puller and pusher swimming microorganisms, as well as direction of the induced flow fields are shown in Fig. \ref{figM1}(a). This simple model has been validated and widely used in the literature \cite{lauga2009hydrodynamics,elgeti2015physics}. In the case of \textit{E. coli} bacteria, for example, the validity of this model has been further confirmed by comparing it to the flow field experimentally measured around an individual swimming cell \cite{drescher2011fluid}.}

\hl{Let us consider a model micro-swimmer swimming toward direction $\bm{e}$, through an unbounded fluid domain. Disturbing flow induced by the swimmer can be modeled as the flow of a force dipole located at instantaneous position of the swimmer ($\xvec_0$). Thrust and drag forces of equal magnitude are exerted in opposite directions ($\pm f_0 \bm{e}$) to the ambient fluid at $\xvec_0 \pm \bm{e} \ l/2$, where the characteristic length $l$ is on the order of swimmer dimensions. For each point force $\fvec$ exerted at point $\xvec_p$ in an infinite fluid domain, the governing equation will turn into:
\ba \label{eqA3}
\nabla P = \eta \ \lap \uvec + \fvec \delta (\xvec-\xvec_p), \quad \nabla \cdot \uvec = 0.
\ea
where $\delta \lp\rvec\rp$ is the Dirac delta function. Equation \eqref{eqA3} can be analytically solved in several ways \cite{chwang1975hydromechanics}, and the resultant velocity field is known as Stokeslet:  
\ba \label{eqA4}
\uvec_S \lp \rvec_p,t \rp = \frac{\fvec}{8\pi \eta} \lp \frac{\bm{I}}{r_p}+\frac{\rvec_p \rvec_p}{r_p^3}\rp \equiv  \bm{G} \cdot \fvec ,
\ea
where $\rvec_p = \xvec -\xvec_p$, and $\bm{G}$ is the corresponding Green's function. A complete set of singularities in Stokes regime, can then be obtained \cite{chwang1975hydromechanics} by taking derivative of the fundamental solution presented in \eqref{eqA4}. The induced flow field of a model force dipole, $\pm f_0 \bm{e}$, located at instantaneous position of the swimmer ($\xvec_0$), can therefore be mathematically expressed as:
\ba \label{eqA5}
\uvec_{\text{SD}} = \frac{\mathscr{D}}{8\pi \eta \ r^3} \lb -1+ 3 \lp \frac{\rvec \cdot \bm{e}}{r} \rp^2 \rb \rvec,
\ea
where $\rvec = \xvec- \xvec_0$, for any generic point $\xvec$ in space. Note that the dipole strength, $\mathscr{D} \approx f_0 l$, has a positive (negative) sign for pusher (puller) swimmers and its value can be inferred from experimental measurements. For instance, the values of $f_0=0.42$ pN and $l=1.9$ $\mu$m, have been experimentally obtained \cite{drescher2011fluid} for E. coli, in agreement with resistive force theory \cite{darnton2007torque}, and optical trap measurements \cite{chattopadhyay2006swimming}.}

Here, we use velocity scale $U_s = f_0/8\pi\eta l$, length scale $L_s = l$, and time scale $T_s = L_s/U_s$, to non-dimensionalize the reported quantities. \hl{Therefore, dimensionless disturbing flow induced by a micro-swimmer reads as:
\ba \label{eqA6}
\bar{\uvec}_{\text{SD}} = \frac{c_0}{\bar{r}^3} \lb -1+ 3 \lp \frac{\bar{\rvec} \cdot \bm{e}}{\bar{r}} \rp^2 \rb \bar{\rvec},
\ea
where $c_0=+1$ ($-1$) for pushers (pullers) and bar signs denote dimensionless quantities.} It worths noting that near-field of the flow induced by a micro-swimmer can be described more accurately via including an appropriately chosen combination of higher order terms from the multipole expansion \cite[e.g.][]{ghose2014irreducible,mathijssen2016upstream}. However, here we are interested in the \textit{span} of swimmers' induced disturbances and their consequent detection region, for which far-field of the flow is of primary interest.

To assess optimality of various swarm arrangements (in stifling disturbing effects), one needs to first quantify the induced fluid disturbances. A measure of distortion (caused by a flock of swimmers to the ambient fluid) can be obtained \cite{mirzakhanloo2019active} by directly computing the Mean Disturbing Flow-magnitude (MDF) over a surrounding ring ($\mathcal{C}$) of radius $R$, i.e. \hl{
\ba
\text{MDF} = \frac{1}{2\pi \bar{R}} \ \oint_{\mathcal{C}} |\bar{\uvec}_{\text{net}}| \text{ds},
\ea	
where $\bar{R} = R/L_s$, and $\bar{\uvec}_{\text{net}}$ is the overall dimensionless flow field induced by the flock. Due to linearity of the Stokes equation \eqref{E1}, the net disturbing flow ($\bar{\uvec}_{\text{net}}$) is computed through superposition of the flow fields ($\bar{\uvec}_{\text{SD}}$) induced by individual swimmers forming the flock (i.e. $\bar{\uvec}_{\text{net}} = \sum_{i = 1}^{N} \bar{\uvec}_{\text{SD}}^i$). Alternatively, one can quantify swarm's induced fluid disturbances by} computing \textit{Area of the Detection Region} (ADR), within which disturbances exceed a predefined threshold. \hl{More precisely, the detection region refers to the subset of space, within which the magnitude of net induced disturbing flow ($\bar{\uvec}_{\text{net}} = \uvec_{\text{net}}/U_s$) exceeds a predefined threshold ($\bar{u}_{\text{th}} = u_{\text{th}}/U_s$), i.e.
\ba
\mathscr{R} = \lcb \forall \xvec : \ |\bar{\uvec}_{\text{net}} \lp \xvec \rp| \geq \ \bar{u}_{\text{th}} \rcb,
\ea	
}which is also consistent with previous numerical studies on swimming microorganisms \cite{doostmohammadi2012low}. The threshold value ($u_{\text{th}}$) can be tuned based on characteristics of the specific problem of interest. For the system representing a prey swarm, as an example, it can be inferred from experimental observations on sensitivity of predators' receptors in sensing flow signatures.


\hl{ \section{Results \& Discussion} }

Here, we combine the described theoretical analysis with direct computations and non-linear optimization, to perform a systematic parametric study on flocks of $N\geq2$ swimmers with a bottom-up approach. Specifically, we develop a general procedure to determine optimal swarming configurations, and systematically investigate their significance in reducing the swarm's induced fluid disturbances. We then present computational evidences demonstrating how such a concealed swarm can actively gather around a favorite spot, point toward a target, or track a desired trajectory, while minimally disturbing the ambient fluid. As a benchmark, here we consider planar arrangements/movements of pusher swimmers (say, \textit{E. coli} bacteria) in an infinite fluid domain. Nevertheless, the reported concealed arrangements will be the same for pullers, and our study can be inherently extended to three-dimensional (3D) scenarios (see Appendix A for details).

\hl{ \subsection{Concealed Arrangements} } \label{sec.arrangement}

Let us consider simple groups of only two and three swimmers. The relative orientation of the swimmers primarily controls the amount of distortion (measured in terms of MDF/ADR) they induce to the surrounding environment. Our results reveal that by swimming in optimal orientations, swimmers can reduce their induced disturbances by more than 50\% (Fig. \ref{figM1}b-c) compared to when they simply swim in schooling orientations (i.e. toward the same direction). In fact, there exist a \textit{range} of optimal swarm configurations, arranging into which will result in minimally disturbing the surrounding fluid. For instance, when two of the agents (in a group of three) swim in directions normal to each other, the swarm arrangement remains optimal regardless of the third one's swimming direction (see the green dashed line in Fig. \ref{figM1}c). This is due to axisymmetric nature of the disturbing flow induced by two perpendicular dipoles (Fig. \ref{figM1}b-II).

It bears attention that computed values of induced disturbances (in terms of ADR/MDF) vary depending on associated hyper-parameters -- specifically, the threshold value ($\bar{u}_{th}$) considered in computing ADR, or the radius ($R/L_s$) of surrounding ring over which MDF is computed. However, such dependencies can be avoided by normalizing the computed values against MDF/ADR induced by a reference swarm arrangement (see e.g. Fig. \ref{figM1}b). The consequent normalized values also represent a cross-match for MDF versus ADR (Fig. \ref{figM1}b). This further confirms equivalence of the described measures in assessing optimality of swarm arrangements.

For groups including a larger number of swimmers (i.e. those with $N>3$ agents), plotting MDF (or ADR) over the entire parameter-space is not practical. However, we know that any flock of $N \in \lcb 4,5, \dots \rcb$ swimmers can be divided into sub-groups of only two or three agents. For each of such sub-groups, the optimal region configurations is then readily available (Fig. \ref{figM1}b-c), assuring about 50\% reduction of induced disturbances. This along with linearity of Stokes equations \eqref{E1}, which describes dynamics of the flow around micro-swimmers, guarantee the existence of optimal swarm arrangements with the same 50\% concealing efficiency. Therefore, the optimal region of configurations exists for any flock of $N$ swimmers, and one can extract an optimal arrangement by implementing non-linear optimization over the parameter-space.

Note that objective functions quantifying swarm's induced disturbances (i.e. MDF/ADR), are nonlinear and often subject to constraints (e.g. minimum separation distance between the agents). Thereby, in search of the global minima by starting from multiple points, here we perform sequential quadratic programming using local gradient-based solvers.\hl{It worths mentioning that merely applying gradient-based solvers will only find local optima, depending on the starting point. To avoid this, here} the starting points are generated using a scatter-search mechanism \cite{glover2003scatter}, which is a high-level heuristic population-based algorithm, designed to intelligently search on the problem domain. Its deterministic approach in combining high-quality and diverse members of the population -- rather than extensive emphasis on randomization -- makes it faster than other similar evolutionary mechanisms, such as genetic algorithm \cite{marti2005scatter}.

\begin{figure}[!htb]
	\centering 
	\includegraphics[width=0.47\textwidth]{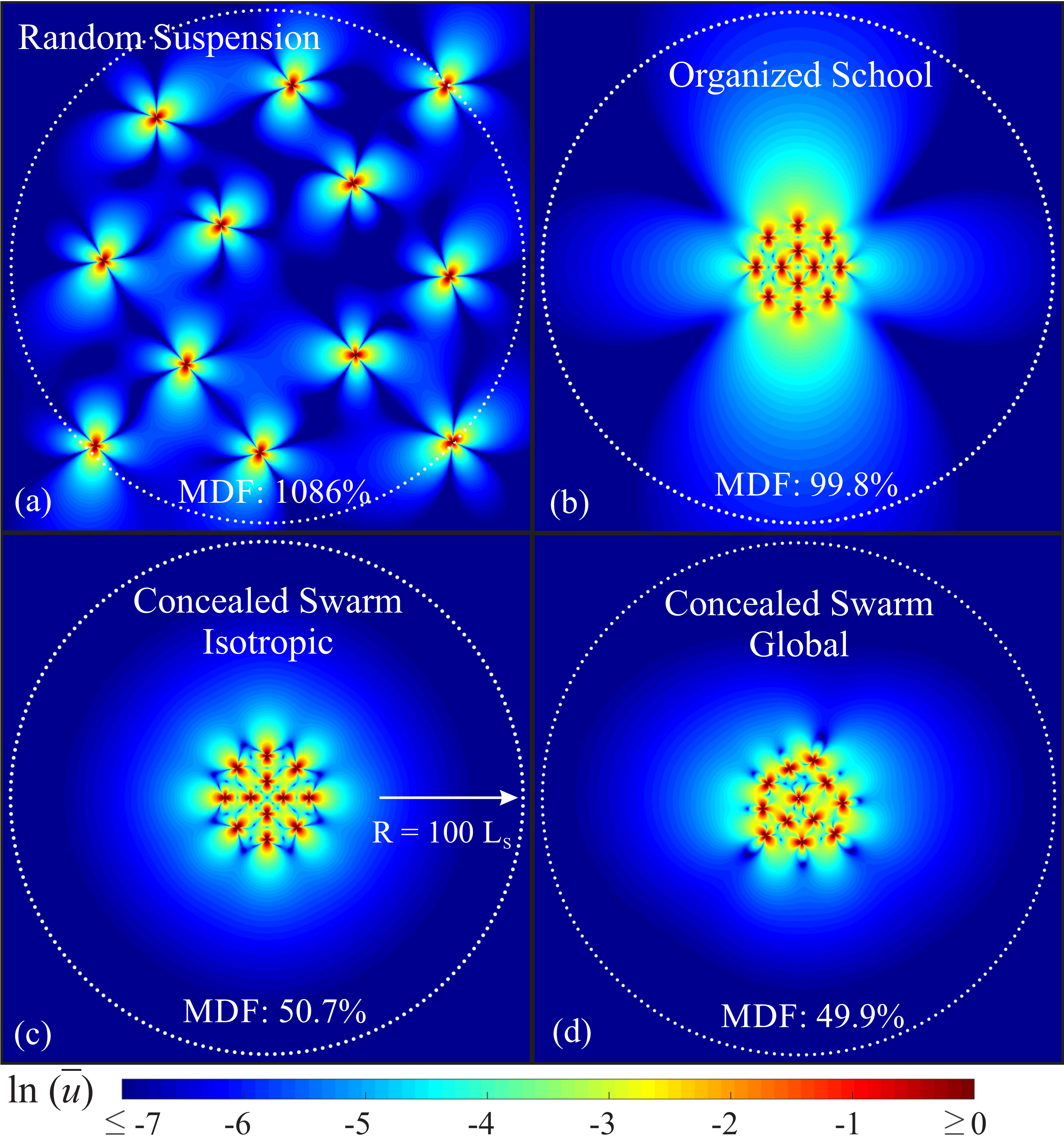}
	\caption {Magnitude of the disturbing flows induced by a randomly arranged suspension of twelve swimmers (a), is compared to when they form a structured group and arrange into an isotropic arrangement with schooling orientation (b), or concealed swarms (c-d). The minimum separation distance ($\xi$) between the swimmers in all cases is set to $10L_s$. Note that positions of the swimmers in panel (b) are the same as those in panel (c). The only difference is that swimmers are arranged into the schooling orientation in (b), whereas the isotropic arrangement in (c) is a particular example from the infinite pool of concealing arrangements (i.e. the optimal region of configurations). For a swarm of twelve swimmers with minimum separation distance of $\xi/L_s = 10$, panel (d) represents arrangement of the globally optimal swarm with minimal disturbing effects. Color shading represents the flow magnitude, and MDF is computed over white dotted rings (with radius $R/L_s = 100$). The reference case used to normalize reported MDF values, corresponds to the case of twelve aligned swimmers all located at the center point.}
	\label{figM2}
\end{figure}

\begin{figure}[!htb]
	\centering 
	\includegraphics[width=0.47\textwidth]{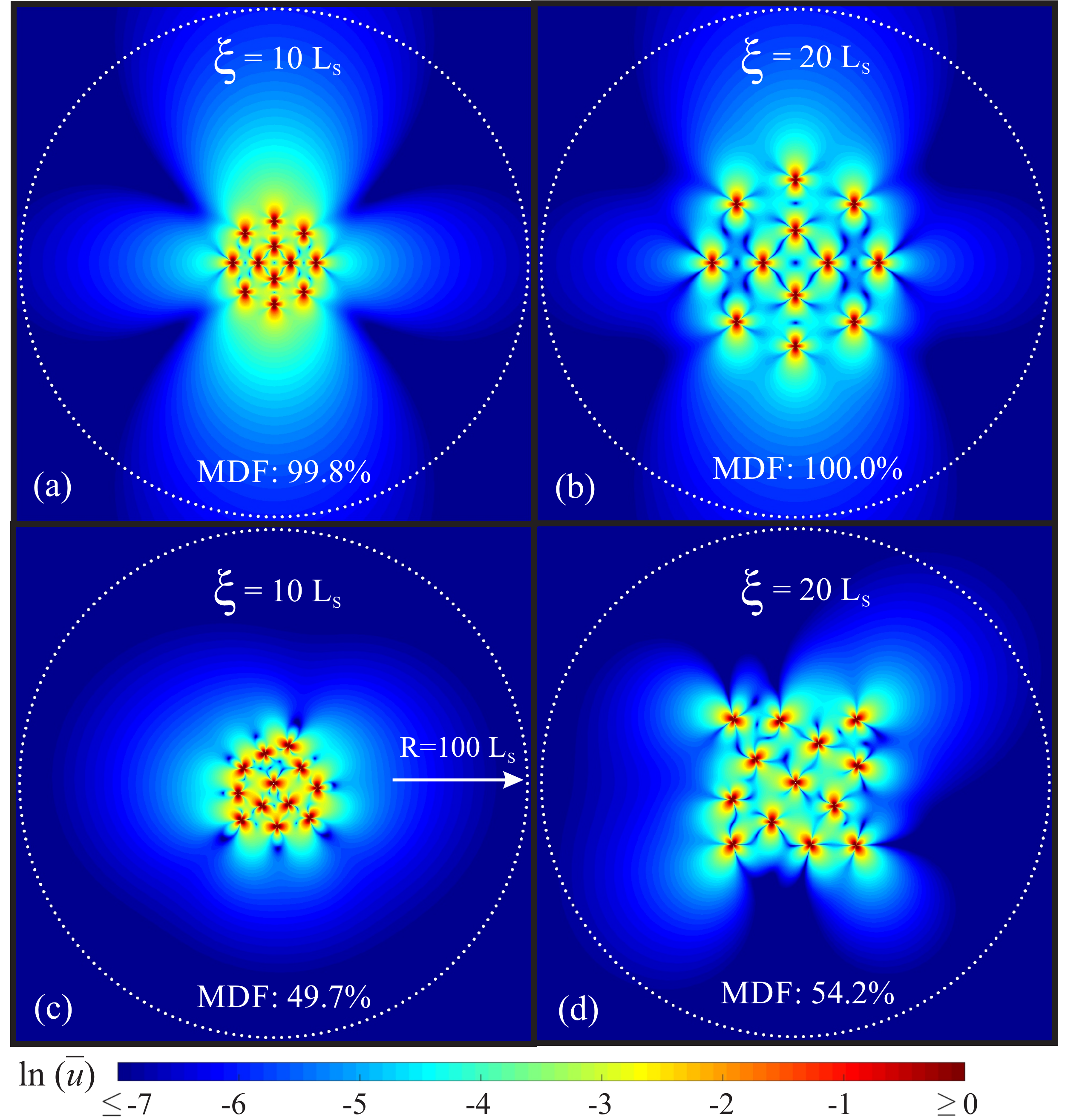}
	\caption {
		\hl{(a)-(b): Magnitude of the net disturbing flows induced by a flock of twelve swimmers schooling in an isotropic positioning with minimum separation distance $\xi/L_s = 10$ and $20$ between them. (c)-(d): Magnitude of the net disturbing flows induced by concealed swarms of twelve swimmers with minimum separation distance $\xi/L_s = 10$ and $20$, respectively. Here the reported values of MDF are computed over the white dotted rings (with radii $R/L_s=100$), and normalized against the reference case. Color shading represents the flow magnitudes in each panel.}
	}
	\label{figS1}
\end{figure}

As a benchmark, magnitude of the disturbing flow induced by a random suspension of twelve swimmers, and the one induced by the same group arranged into an isotropic organized school are compared in Fig. \ref{figM2}, to that induced by concealed swarms of twelve swimmers. It worth noting that the amount of disturbances induced by a flock of swimmers depends also on the minimum separation distance between the agents. Our numerical results show that the role of this factor is more significant for a concealed swarm than for an organized school \hl{(see e.g. Fig. \ref{figS1})}. However, it can still be considered as a minor factor compared to relative orientation of the swimmers.

%


\hl{ \subsection{Concealed Swarming} } \label{sec.swarming}

There exist many situations (for both motile microorganisms and swimming micro-robots) that swimmers form an active swarm (i.e. a disordered cohesive gathering) around a desired spot. For instance, this can be a swarm of bacteria around a nutrient source \cite{kearns2010field}, or a flock of biological micro-robots performing a localized micro-surgery \cite{nelson2010microrobots,hu2018soft}. Remaining concealed in these scenarios can keep a bacterial swarm stealth from nearby predators, or help keeping a deployed flock of micro-robots non-disturbing to the host medium.

\begin{figure}[!htb]
	\centering 
	\includegraphics[width=0.48\textwidth]{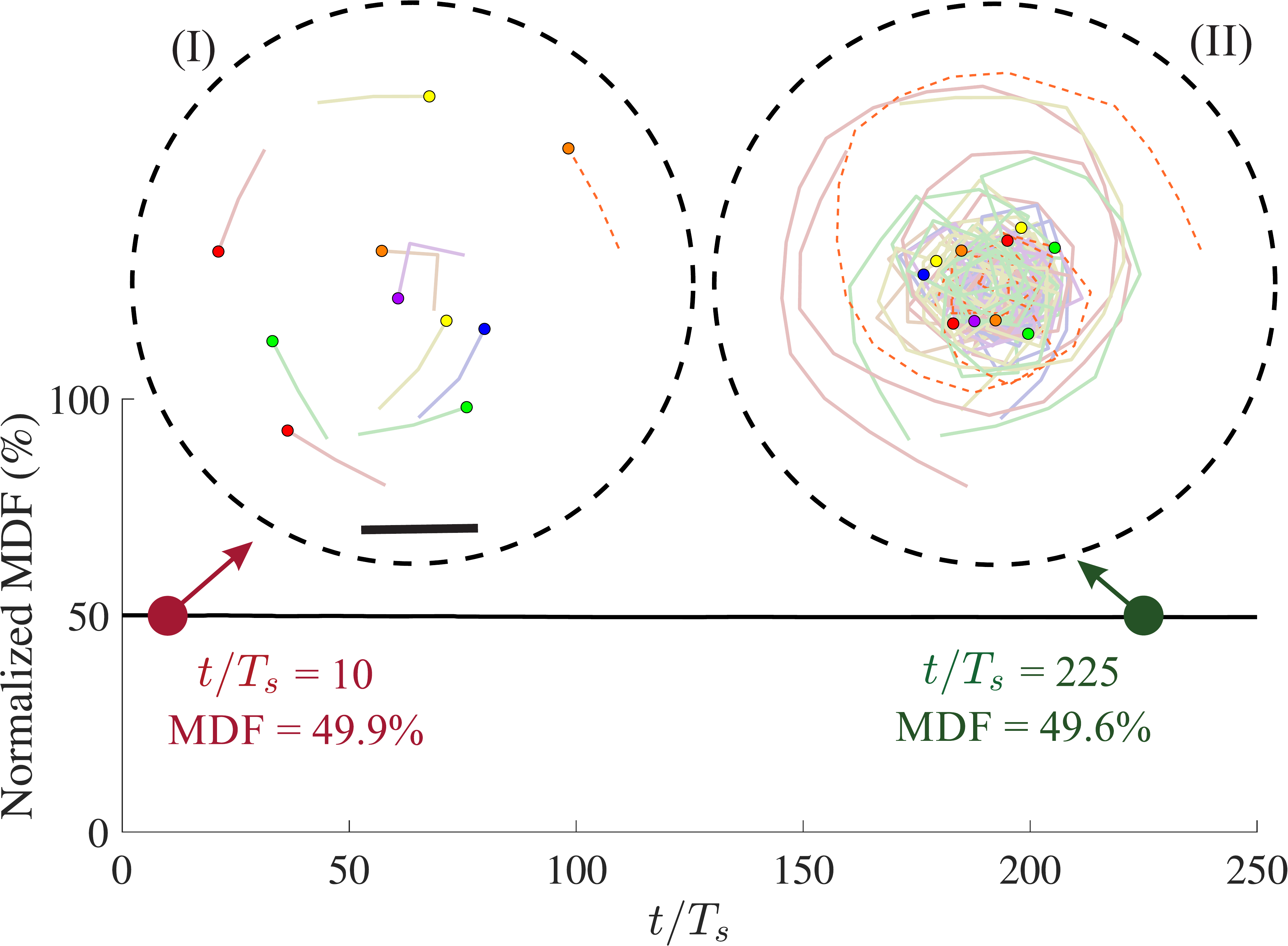}
	\caption {Fluid disturbances (in terms of MDF) monitored over the time evolution of an \textit{active} concealed swarm of ten swimmers (Movie S1). The agents are initially positioned and oriented randomly (at $t/T_s = 0$). Thereafter, each agent represents a version of run-and-tumble dynamics (with $\tau_r/T_s = 5$), so that to keep the swarm's arrangement within the optimal region of configurations at every instant of time. The insets (I and II) represent snapshots of the swarm's activity at an early ($t/T_s = 10$) and late ($t/T_s = 225$) stages of the system evolution, respectively. The immediate values of MDF (computed over surrounding rings of radii $R/L_s = 100$) for the presented snapshots are $49.9\%$ and $49.6\%$, respectively. In each panel, the instantaneous positions of the swimmers are marked with colored dots, and their trajectories are shown by gray-scale lines. As a benchmark, we have also highlighted one of the trajectories with an orange dashed-line. The presented scale bar is $10L_s$.}
	\label{figM3}
\end{figure}

Note that forming a swarm (as opposed to a random suspension), by itself, keeps the net induced distortions bounded (see e.g. Fig. \ref{figM2}). However, to have \textit{minimal} disturbing effects, arrangement of the swimmers (forming a swarm) must lie within the optimal region of configurations at every instant of time. To this end, we choose the proposed measure of net induced disturbances (i.e. MDF) as the objective function ($\mathcal{Z}$) to be minimized by the swarm arrangement. \hl{Dynamics of an active \textit{concealed} swarm can then be described as follows: (i) swimmers forming the swarm get into an optimal arrangement (with minimal fluid disturbances); (ii) each swimmer then swims steadily forward (i.e. `runs') for a fixed period of time (say, $\tau_r$); (iii) the swimmers will then reorient quickly (i.e. `tumble') into a new optimal arrangement, and then (iv) start running once again toward the new directions. This sequence of events occur in turn, repeatedly. Parameters including swimming speed and frequency of tumbling events ($\tau_r^{-1}$) can be tuned according to the system of interest.} As a benchmark, here we present a sample time evolution of an active concealed swarm in Fig. \ref{figM3}. Through the above described dynamics, the swarm remains cohesive, keeps itself confined within a finite region of space (around a desired spot), and is able to stifle the induced disturbances by $\sim 50\%$ through system evolution (Fig. \ref{figM3}). This is equivalent to shrinking the swarm's detection region by half.

\hl{It also worths mentioning that motion of each individual swimmer in the presented system, can be seen as a \textit{controlled} version of the so-called \textit{run-and-tumble} mechanism. This is inspired by the observed behavior of swimming microorganisms, such as E. coli bacteria -- known as the paradigm of run-and-tumble locomotion \cite{berg2008coli}. Recent observations \cite{polin2009chlamydomonas} reveal that even C. reinhardtii cells swim in a version of run-and-tumble. From a practical point of view, realization of the smart form of run-and-tumble mechanism also seems feasible in the context of internally/externally controlled artificial micro-swimmers. The recently proposed \textit{Quadroar} swimmer \cite{jalali2014versatile,mirzakhanloo2018flow}, for instance, propels (i.e. runs) on straight lines, and can perform full three-dimensional (3D) reorientation (i.e. tumbling) maneuvers \cite{saadat2019experimental}.}

\hl{ \subsection{Stealthy Maneuvers: \\ Target Pointing and Trajectory Tracking} } \label{sec.target}

Through altruistic collaborations, micro-swimmers can also remain stealth while traveling toward a target point or tracking a desired trajectory in space. There is only one caveat here. The objective function ($\mathcal{Z}$), to be minimized by the traveling swarm during each consecutive run, must now represent a measure not only for the overall disturbances induced by the swimmers, but also their distances from the target point (or from the desired trajectory). Therefore, we define
\ba \label{Eq5}
\mathcal{Z}= \epsilon \times \bar{\text{MDF}} + (1- \epsilon) \times \bar{\text{RMS}}_d \ ,
\ea 
where $\bar{\text{RMS}}_d$ stands for the normalized root mean square of swimmers' distances from the target point; $\bar{\text{MDF}}$ quantifies the overall induced disturbances; and $0 \leq \epsilon \leq 1$ is the detuning parameter which determines the importance of concealing versus travel time. Note that the ratio $\epsilon/(1- \epsilon)$ properly covers the entire span of $[0,\infty)$ when $\epsilon \in [0,1)$. In practice, a variety of imaging techniques \cite{pane2019imaging} can be used to feed back the state information (including swimmers' distance toward the target point) into the agents' decision-making unit. However, some bio-hybrid systems may rely on local (point-wise) sensing capabilities of swimming microorganisms (in measuring quantities such as light or chemicals) to obtain state information. To optimally control such systems, a more realistic objective function can be devised (\hl{Appendix B}) by replacing $\bar{\text{RMS}}_d$ in Eq. \eqref{Eq5} with $\bar{\text{RMS}}_\theta$. The latter stands for normalized root mean square of the swimmers' deviations from their locally desired directions, in the absence of concealing interests. Nevertheless, our numerical experiments reveal that the bottom-line of analysis conducted using such an alternative objective function still remains the same (\hl{Fig.\ref{figS4}, Appendix B}).

\begin{figure}[!htb]
	\centering 
	\includegraphics[width=0.49\textwidth]{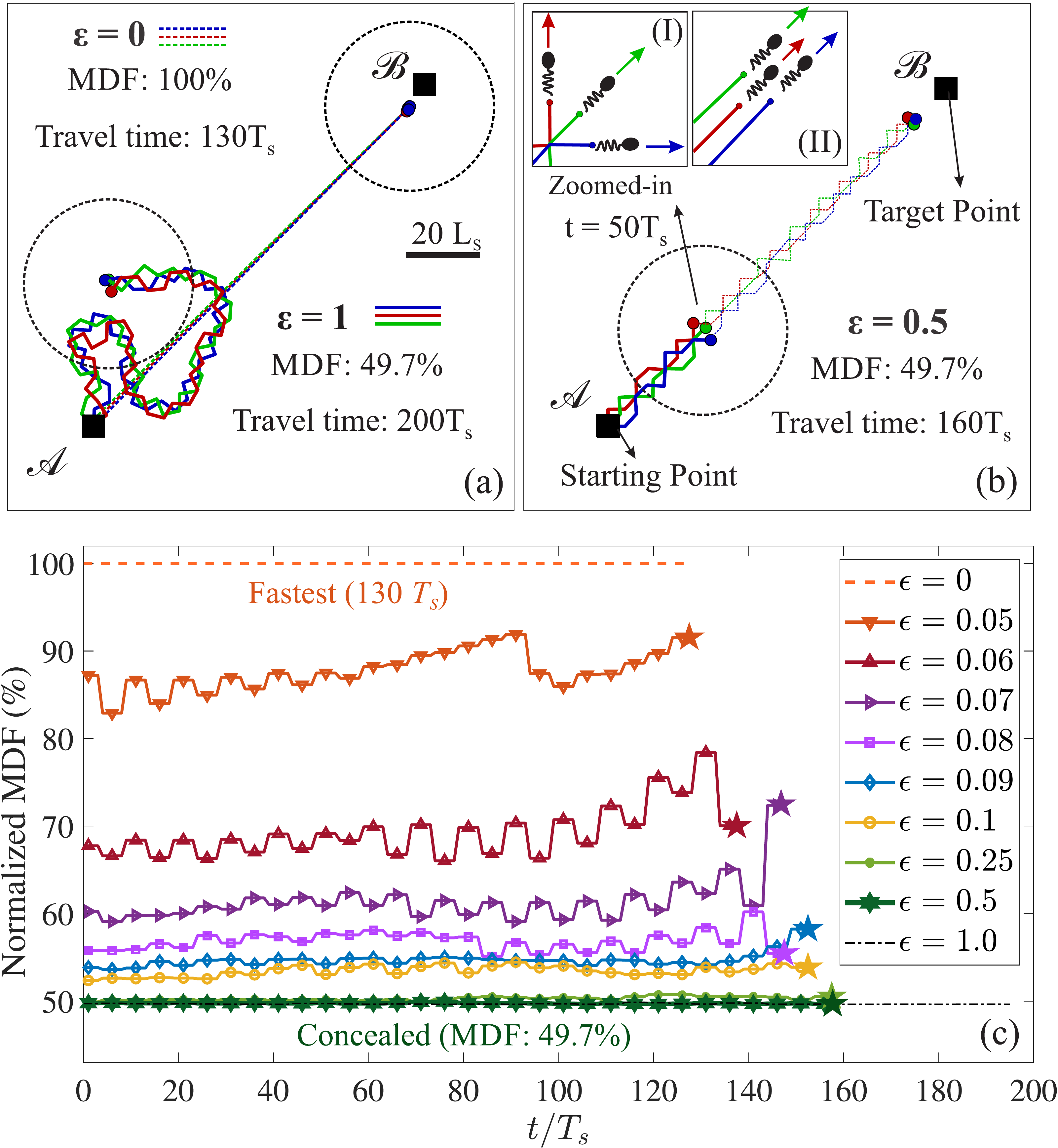}
	\caption {
		Sample flocks of micro-swimmers, controlled by different values of $\epsilon$, so as to travel from a starting point ($\mathscr{A}$) toward the target point (at $\mathscr{B}$). \textbf{(a):} Comparison between trajectories of the swimmers when controlled by extreme values of $\epsilon$ = 0 and 1, respectively. Denoted by dashed thin lines are trajectories of the fastest traveling swarm (travel time = $130T_s$) with no concealing ($\epsilon = 0$). On the other hand, trajectories of the traveling swarm with the highest possible concealing efficiency (MDF = $49.7\%$), yet no constraints on preferred direction ($\epsilon = 1$), are denoted by solid thick lines. The latter never reaches the target point (Movie S2). \textbf{(b):} Snapshots of the optimal concealed traveling swarm ($\epsilon = 0.5$) having the most possible concealing efficiency in the cost of only 23\% increase in the travel time (Movie S3). Blue, green, and red thick (thin) solid (dashed) lines represent trajectories of the swimmers after $t/T_s = 50$ ($160$). Instantaneous arrangement of the swimmers at $t/T_s=50$ is schematically shown in a magnified view as inset (I) and is compared to the arrangement of an organized school (II) with MDF of about 100\%. \hl{The dashed circles in panels (a) and (b) mark instantaneous position of each flock in the presented snapshots.}
		\textbf{(c):} Fluid disturbances induced by the traveling swarms controlled with various values of $\epsilon$, are measured in terms of MDF and monitored during their trip from $\mathscr{A}$ to $\mathscr{B}$. The terminal time, at which a swarm reaches to the target point $\mathscr{B}$, is denoted in each case by an asterisk. 
		}
	\label{figM4}
\end{figure}

Sample flocks of micro-swimmers, controlled to travel from a starting point ($\mathscr{A}$) toward a target point ($\mathscr{B}$) in \textit{stealth} versus \textit{fast} modes (tuned by $\epsilon$), are shown in Fig. \ref{figM4}. Note that $\epsilon=0$ corresponds to the fastest traveling swarm, for which the swimmers travel in schooling arrangements (i.e. toward the same direction), but it provides no concealing benefits (i.e. MDF = $100\%$ for $\epsilon=0$). On the other extreme, i.e. for $\epsilon=1$, the swarm will have the highest concealing efficiency (MDF = $49.7\%$), yet never reaches the target point (Fig. \ref{figM4}a and Movie S2). Trade-off between the travel time and the overall efficiency of concealing is demonstrated with more details in Fig. \ref{figM4}(c). To illustrate, we have monitored the induced fluid disturbances for traveling swarms controlled with various values of $\epsilon$, during their migration from $\mathscr{A}$ to $\mathscr{B}$. As $\epsilon \rightarrow 1$ ($\rightarrow 0$), the swarm will travel slower (faster), yet induces less (more) disturbances to the ambient fluid.

\hl{Recall that for a traveling swarm controlled by $\epsilon=0$, the objective function \eqref{Eq5} encodes only a measure of the swimmers' distances to the target point. This results in reaching the target point in minimum amount of time (Fig. \ref{figM4}a). However, once an $\epsilon > 0$ is introduced to the swarm control strategy, the objective function (to be minimized by the traveling swarm) will also include a measure for the swimmers' overall induced disturbances (in terms of MDF). Thereby, the fluid disturbance (measured in terms of normalized MDF) induced by a controlled traveling swarm, rapidly decays with $\epsilon$ and eventually converges to the highest possible concealing efficiency (i.e. MDF = $49.7\%$) at $\epsilon \approx 0.5$ (Fig. \ref{figM4}c).} It is remarkable that swarming in such an optimally concealed mode (Fig. \ref{figM4}b and Movie S3), that is the fastest among those with highest possible concealing efficiency, costs only $23\%$ increase in the trip duration compared to the fastest possible swarm. The associated concealing efficiency for such an optimal traveling swarm is equivalent to $\sim$ 50\% shrink of the swarm's detection region throughout its migration from $\mathscr{A}$ to $\mathscr{B}$.

In the end, it also worths noting that although the focus of our analysis in this article has been on planar (2D) swarm arrangements/movements, the present study can be readily extended to three-dimensional (3D) scenarios. As a benchmark, we discuss 3D concealed arrangements in \hl{Appendix A} (\hl{Fig. \ref{figS5}}), for which reduction of the swarm's induced disturbances exceed $99\%$. We then demonstrate how such swarm configurations can help a group of swimmers to remain stealth (with $\geq 90\%$ reduction in MDF) throughout their trip from $\mathscr{A}$ to $\mathscr{B}$ in a 3D space \hl{(Fig. \ref{figS6})}. Additionally, a sample concealed swarm of micro-swimmers tracking a desired trajectory through a \textit{non-uniform} environment is also discussed in \hl{Appendix C} (\hl{Fig. \ref{figS3}} and Movie S4).

\hl{ \section{Concluding Remarks} }

In conclusion, here we revealed that micro-swimmers can form a stealth swarm through controlled cooperation in suppressing one another's disturbing flows. Specifically, our results unveil the existence \textit{concealed} arrangements, which can stifle the swarm's hydrodynamic signature (and thus shrink its detection region) by more than 50\% (or 99\%) for planar (or 3D) movements. We then demonstrated how such a concealed swarm can actively gather around a desired spot, point toward a target, or track a prescribed trajectory in space. Our study provides a road map to optimally control/lead a swarm of interacting micro-robots in stealth versus fast modes. This, in turn, paves the path for non-invasive intrusion of swimming micro-robots with a broad range of biomedical applications \cite{nelson2010microrobots}.

The presented findings also provide insights into dynamics of prey-predator systems. \hl{Importance of the fluid mechanical signals (i.e. flow signatures) produced by swimming objects, in dynamics of prey-predator systems is well appreciated for a broad range of aquatic organisms. Free-living copepods, for example, possess highly sensitive fluid-mechanoreceptors \cite{yen1992mechanoreception} capable of detecting disturbing flows as small as 20 $\mu$ms$^{-1}$. These sensors enable the organism to accurately measure fluid disturbances induced by nearby predators (preys), estimate their distance/size, and thereby properly trigger escape (catch) behavior \cite{fields1997escape,fields2002fluid}. Another example is the Gram-negative \textit{Bdellovibrio bacteriovorus} \cite{rendulic2004predator}, which is a prototypical predator among motile microorganisms and hunts other bacteria, such as \textit{E. coli}. Recent experiments \cite{jashnsaz2017hydrodynamic} show that it is, in fact, hydrodynamics rather than chemical clues that lead this predator into regions with high density of prey. Therefore, quenching the flow signature (and thus shrinking the associated detection region) by swarming in concealed modes, can potentially have a significant impact on trophic transfer rates among a broad range of aquatic organisms.} In particular, stifling the induced disturbances may help an active swarm of prey swimmers gathered around a favorite spot (say, a nutrient source) to lower their detectability (and thus predation risk) through shrinking their detection region. Quenching flow signatures induced by a traveling swarm, on the other hand, may help a swarm of predators to remain concealed while attacking a target prey flock.

\hl{ \section*{Supplementary Material} }

See supplementary material for the Movies S1-S4. 
\\ A complete description of these supplementary movies are provided in Appendix E.

\hl{ \section*{Acknowledgments} }

This work is supported by the National Science Foundation grant CMMI-1562871. Authors would like to thank Dr. Mir Abbas Jalali for valuable discussions.

\hl{ \section*{Data Availability} }

The data that supports the findings of this study are available within the article and its supplementary material.

\vspace{20mm}
\noindent\rule{0.4\textwidth}{0.4pt}
\appendix      

\hl{ \section{Concealed Swarms with \\ Three-dimensional (3D) Arrangements and Movements} } 

\hl{The focus of this article has been on planar (2D) swarm arrangements/movements. However, the present study can be readily extended to three-dimensional (3D) scenarios. In particular, the same procedure can be used to also find 3D optimal (i.e. concealed) swarm arrangements for any flock of $N$ swimmers. The exception is that to quantify fluid disturbances induced by a 3D swarm arrangement, one needs to either: (i) compute the mean disturbing flow-magnitude (MDF) over the surface of a surrounding \textit{sphere}, or (ii) compute the \textit{volume} of a swarm's detection region (VDR).}

\begin{figure}[!htb]
	\centering 
	\includegraphics[width=0.45\textwidth]{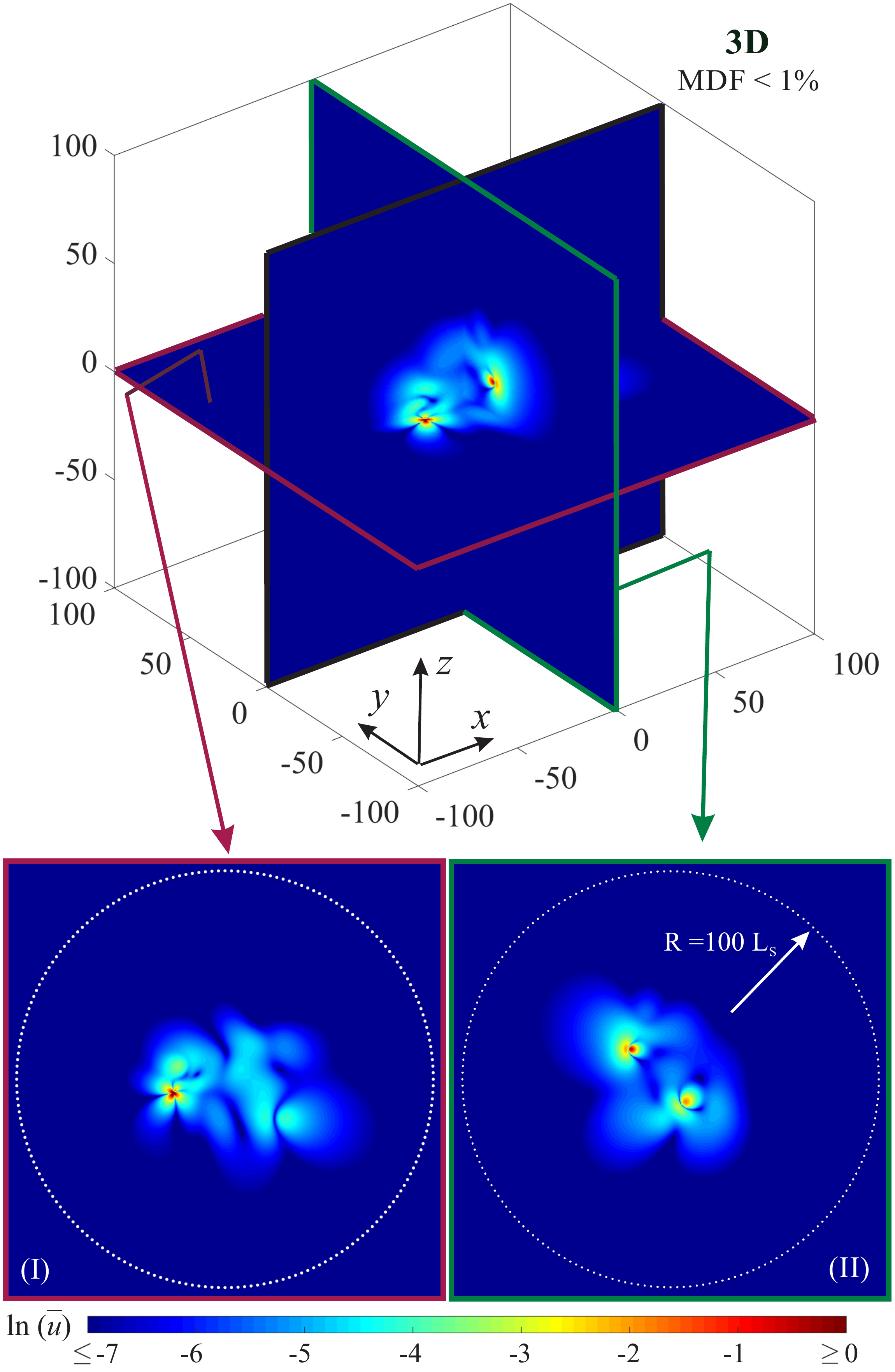}
	\caption {
		\hl{Magnitude of the net disturbing flows induced by a three-dimensional (3D) concealed swarm of $N=12$ swimmers (c.f. the 2D arrangements presented in Fig. \ref{figM2}). Forming the presented optimal swarm arrangement suppresses the induced disturbances (and thus shrinks the swarm's detection region) by more than $99\%$. Here the induced fluid disturbances are quantified by computing mean disturbing flow-magnitude (MDF) over the surface of a surrounding sphere (with $R/L_s=100$). Also, as a benchmark, the minimum separation distance ($\xi$) between swimmers forming the swarm is set to $10 L_s$, which is the same as those presented in Fig. \ref{figM2}. Insets (I-II) represent (x-y) and (y-z) cross sections of the swarm arrangement, respectively. In the panels, color shading represents the flow magnitude, and white dotted rings represent cross-sections of the surrounding sphere used to compute MDF. The reference case used to normalize MDF, corresponds to the case of twelve aligned swimmers (i.e. in schooling arrangement) all located at the center point.}
		}
	\label{figS5}
\end{figure}

\hl{As a benchmark, a 3D concealed arrangement is demonstrated in Fig. \ref{figS5} for the same flock of twelve swimmers presented in Fig. 2 of the article. We highlight that forming such a 3D concealed swarm suppresses the induced disturbance by more than $99\%$ -- compare this value to the $\sim 50\%$ reduction achieved for 2D concealed arrangements of the same flock (Fig. \ref{figM2}). This is equivalent to more than $99\%$ shrink of the instantaneous detection region for the swarm. Note that such a dramatic suppression of the induced disturbances, achieved by forming a 3D concealed swarm (versus a 2D one), reflects a sudden drop in leading order of the swarm's induced disturbing flows. In fact, for any sub-group of three orthogonally oriented agents, within a 3D swarm, the leading order of induced disturbances switches from being a dipole (decaying as $1/r^2$) to a quadrapole (vanishing as $1/r^3$).}

\begin{figure}[!htb]
	\centering 
	\includegraphics[width=0.47\textwidth]{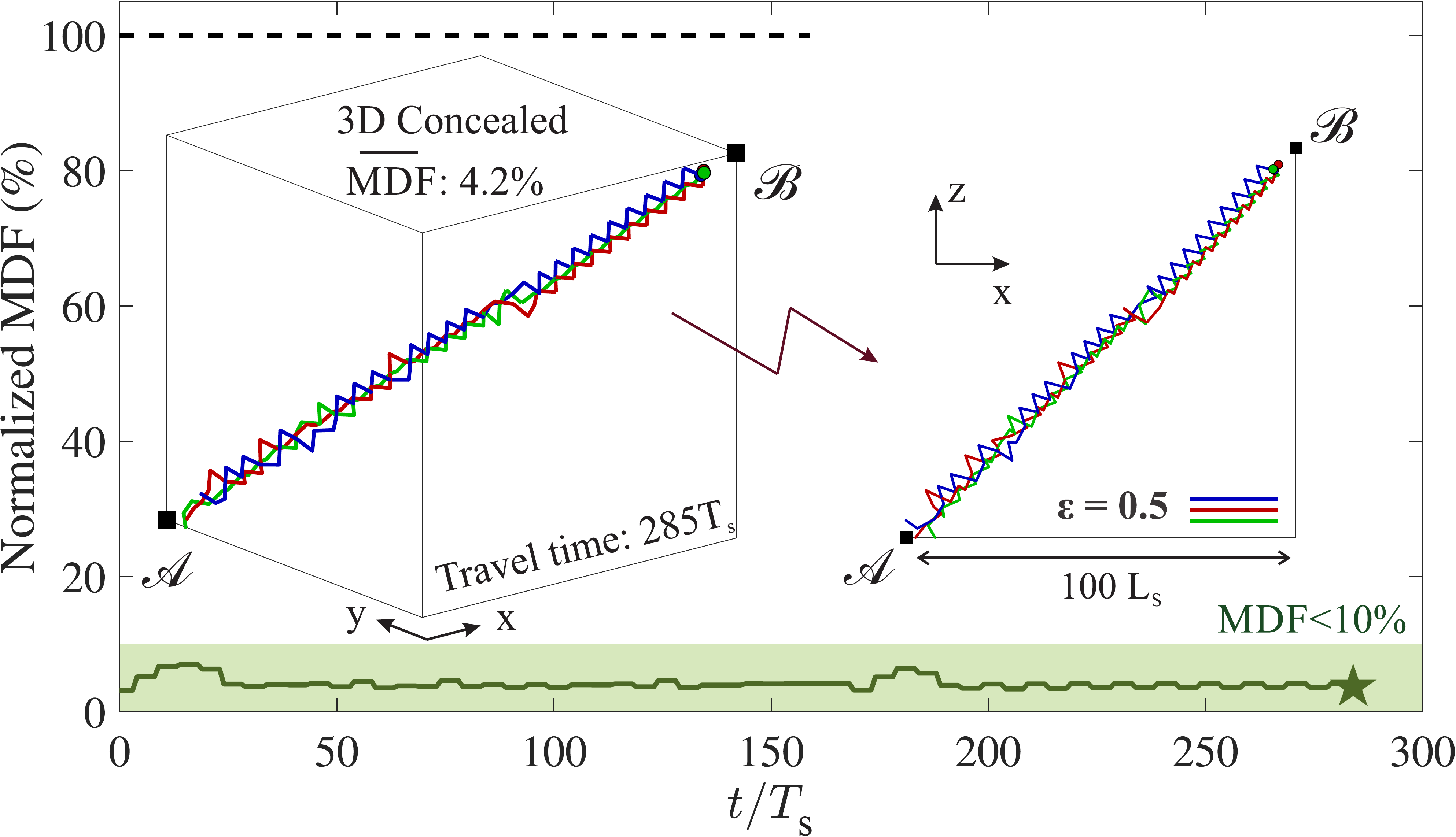}
	\caption {
		\hl{Fluid disturbances induced by a 3D concealed swarm of three micro-swimmers are measured in terms of MDF within an unbounded three-dimensional space, and monitored throughout their migration from a starting point ($\mathscr{A}$) toward a target point ($\mathscr{B}$). The 3D MDF is computed over the surface of a surrounding sphere of radius $R/L_s =100$, and the swarm is controlled by $\epsilon = 0.5$ --see Eq.\eqref{Eq5}. The terminal time, at which the swarm reaches the desired target point at $\mathscr{B}$, is denoted by an asterisk. The insets represent trajectories of the swimmers in full 3D and 2D views, respectively. For the sake of comparison, disturbances induced by the fastest swarm with no concealing efficiency ($\epsilon=0$) is also monitored (in terms of normalized MDF) over time, and is shown by a black dashed line.}
	}
	\label{figS6}
\end{figure}

\hl{Through altruistic collaborations, micro-swimmers also can form a 3D concealed swarm while traveling toward a target point in a 3D space (see Fig. \ref{figS6}). The objective function (to be minimized by the swarm through cooperation of the agents) remains untouched, and one can follow the same procedure (as outlined in the paper) to find the optimally concealed traveling swarm. The caveat here is that the 3D MDF is computed over the surface of a surrounding \textit{sphere}. As a benchmark, here we show (in Fig. \ref{figS6}) a sample 3D concealed swarm of micro-swimmers traveling from a starting point ($\mathscr{A}$) toward a target point (at $\mathscr{B}$). Our numerical experiments reveal that by setting $\epsilon = 0.5$ the traveling swarm can shrink its detection region by more than $90\%$ and remain concealed throughout the trip from $\mathscr{A}$ to $\mathscr{B}$ (see Fig. \ref{figS6}).}

%

\hl{ \section{\\ Stealth Target Pointing via Local Sensory Information} }

\begin{figure}[!htb]
	\centering 
	\includegraphics[width=0.46\textwidth]{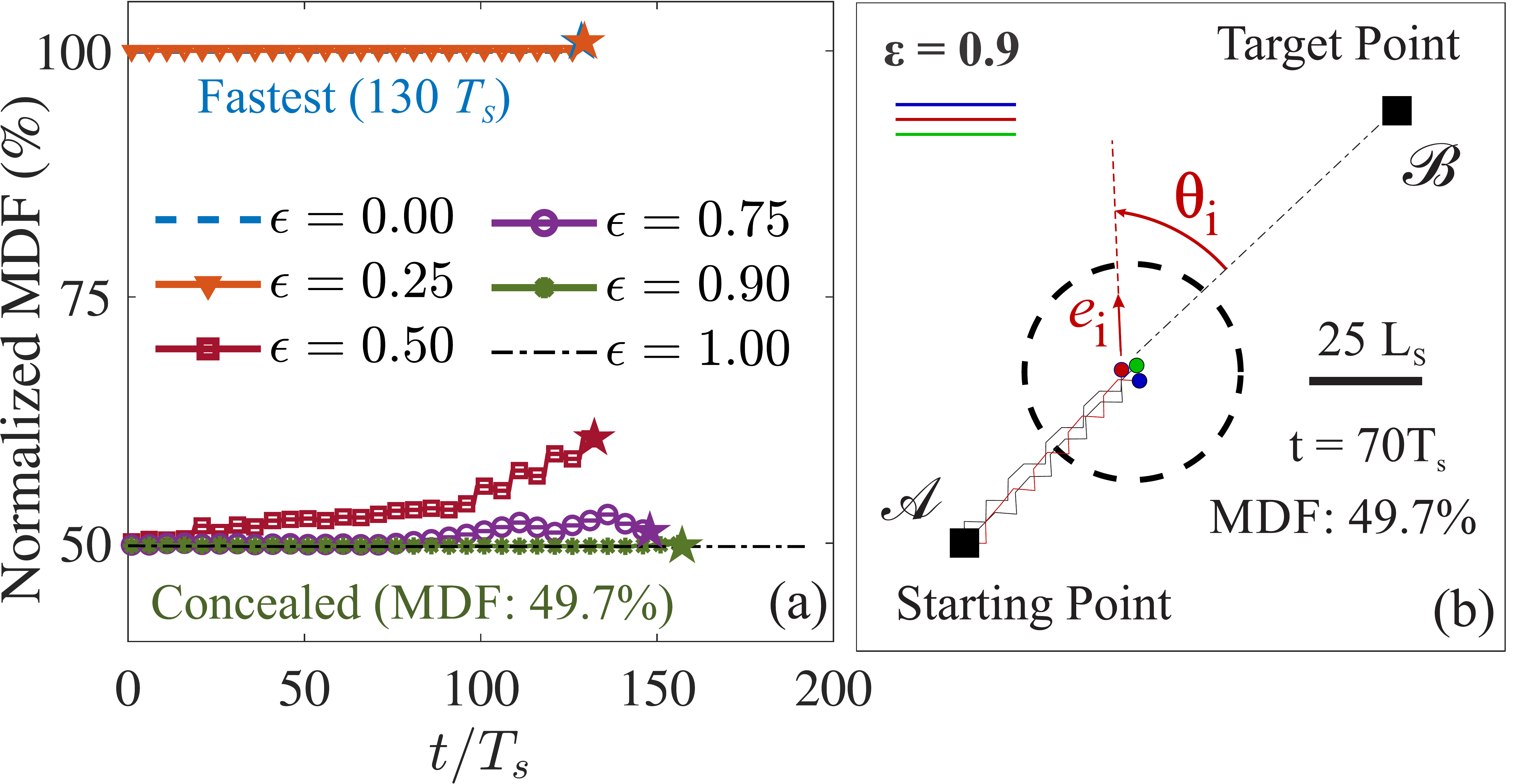}
	\caption {
		\hl{(a) Fluid disturbances induced by sample flocks of micro-swimmers, which are controlled to migrate from a starting point ($\mathscr{A}$) toward a target point ($\mathscr{B}$) with different values of $\epsilon$ tuning their alternative objective function \eqref{Eq5A}. The presented traveling swarms can represent flocks of micro-swimmers swarming from $\mathscr{A}$ to $\mathscr{B}$ under influence of an external gradient, the intensity of which being tuned by different values of $1-\epsilon$. The terminal time, at which a swarm reaches the target point at $\mathscr{B}$, is denoted in each case by an asterisk. Note that the swarm corresponding to the extreme case of $\epsilon=1$, never reaches the target. (b) Snapshots of the optimally concealed traveling swarm (controlled with $\epsilon = 0.9$) which has the highest possible concealing efficiency (MDF = 49.7\%) in cost of only 23\% increase in the travel time compared to the fastest possible swarm. As a benchmark, the deviation angle ($\theta_i$) for the swimmer $i$ swimming in direction $e_i$ is shown (in red). Blue, green, and red solid lines represent trajectories of the swimmers after $70T_s$.}
	}
	\label{figS4}
\end{figure}

\hl{In the presented study we demonstrated that through altruistic collaborations, micro-swimmers can form a concealed swarm while traveling toward a target point or track a desired trajectory in space. These traveling swarms can represent: (i) flocks of swimming micro-robots traveling (in vivo) toward a target point while controlled to be fast/concealed (as tuned by $\epsilon$); (ii) a swarm of predators attacking a target prey flock (at point $\mathscr{B}$) in stealth versus fast modes (tuned by $\epsilon$); or even (iii) flocks of motile microorganisms swarming under influence of an external gradient (the intensity of which being modeled as $1-\epsilon$) from $\mathscr{A}$ to $\mathscr{B}$ -- e.g. in chemotaxis of sperm cells toward an egg. However, we note that sensing capabilities of swimming microorganisms are often limited to the point-wise measurement of various quantities -- e.g. light or chemicals \cite{adler1975chemotaxis}. This makes them unable to identify their distance toward a desired target point, as required for the model presented in Eq. \ref{Eq5} of the article. Therefore, for a bio-hybrid system which relies on sensing capabilities of swimming microorganisms, a more realistic objective function ($\mathcal{Z}$) can be devised by replacing $\bar{\text{RMS}}_d$ in Eq. \ref{Eq5} with $\bar{\text{RMS}}_\theta$, which stands for the normalized root mean square of the swimmers' deviations from their locally desired directions in the absence of concealing interests. The alternative objective function thereby reads as
\ba \label{Eq5A}
\mathcal{Z}= \epsilon \times \bar{\text{MDF}} + (1- \epsilon) \times \bar{\text{RMS}}_\theta \ .
\ea 
Note that when concealing is not of interest, the desired direction at the location of each swimmer is the steepest ascent in the external field that leads it toward the target. Thus, $\bar{\text{RMS}}_\theta$ stands for the normalized root mean square of the swimmers' deviations from the direction representing maximal gradient -- i.e. toward the target (see Fig. \ref{figS4}b). Mathematically, we define
\ba
\bar{\text{RMS}}_\theta = \sqrt{ \lp \sum_{i=1}^{N} \theta_i^2 \rp/N } \ .
\ea 
This requires the swimmers to only identify deviation of their swimming direction from that of the maximal gradient in the external field of interest, that can be obtained locally.}

\hl{Trade-off between the travel time and the overall efficiency of concealing in this case is demonstrated with more details in Fig. \ref{figS4}. The induced fluid disturbances (measured in terms of MDF) are monitored during the trip from $\mathscr{A}$ to $\mathscr{B}$, while the traveling swarm is controlled with various values of $\epsilon$. Similar to what was observed for the test cases presented in Fig. \ref{figM4}, as $\epsilon \rightarrow 0$ ($\rightarrow 1$) the swarm travels faster (slower) in space, i.e. the travel time decreases (increases), but it will induce more (less) disturbances to the ambient fluid. Our results reveal that the bottom-line of our analysis also remains the same. In particular, the results reveal that swarming in an optimally concealed mode (via such an alternative local sensory information), with more than $50\%$ reduction in disturbances, may cost only $23\%$ increase in the trip duration compared to the fastest possible trip (Fig. \ref{figS4}). This is equivalent to 50\% shrink in detection region of the swarm throughout its migration from $\mathscr{A}$ to $\mathscr{B}$. }

\hl{ \section{\\ Stealth Trajectory Tracking in a non-Uniform Environment} }

\hl{There exist many situations, for both biological micro-robots and swimming microorganisms, in which they have to travel through a non-uniform environment. Examples include fluids at the interface of different organs inside the human body with distinct viscosities, or those in vicinity of a mucus zone \cite{liebchen2018viscotaxis}. Depending on their propulsion mechanism, motile microorganisms experience different energy expenditures, and thus distinct swimming speeds, while traveling in regions with different rheological properties \cite{kaiser1975enhanced,berg1979movement,shen2011undulatory,liu2011force,spagnolie2013locomotion,martinez2014flagellated}.}

\begin{figure}[!htb]
	\centering 
	\includegraphics[width=0.49\textwidth]{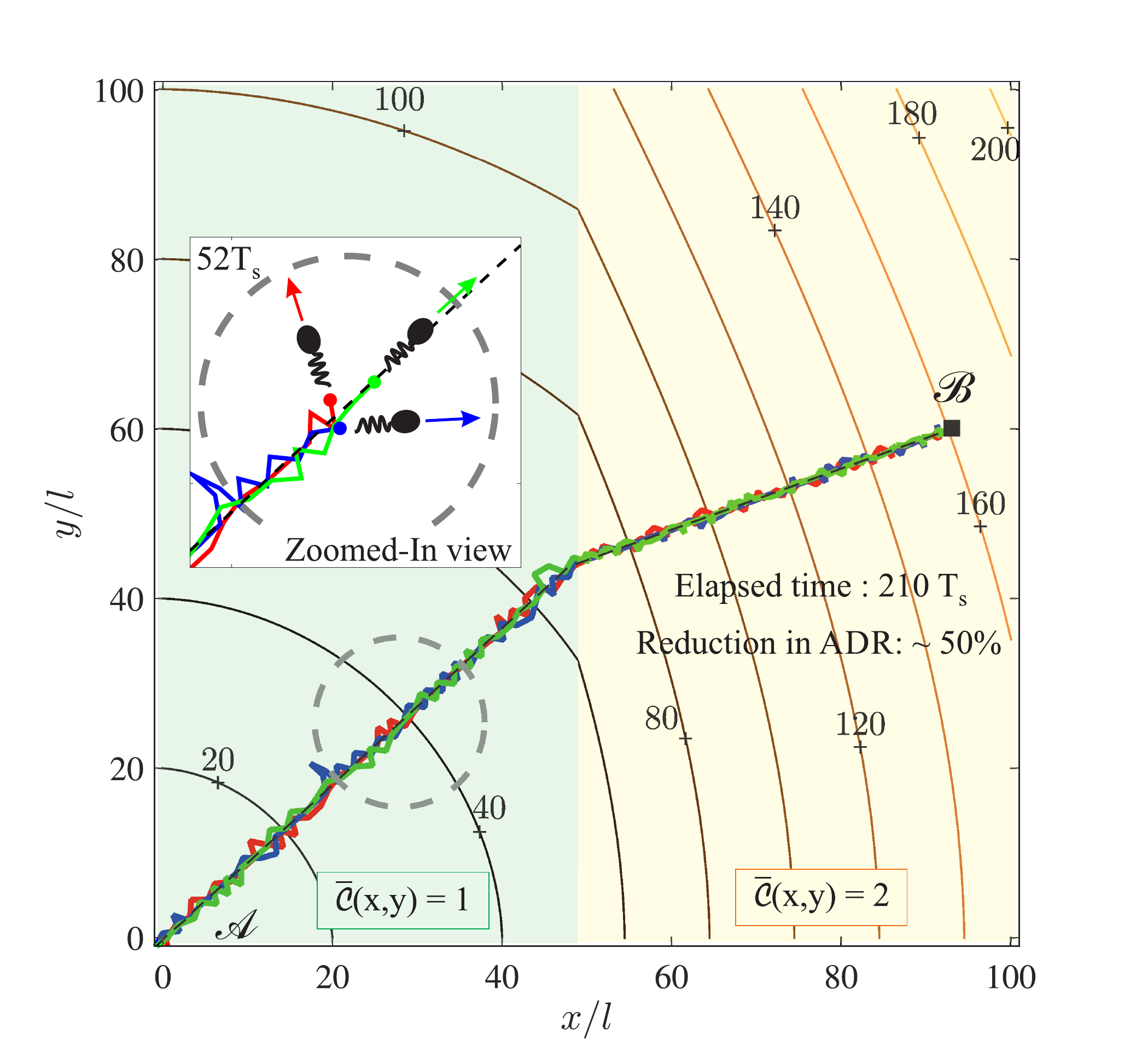}
	\caption {
		\hl{A concealed swarm of three swimmers tracking the optimal trajectory in a non-uniform environment from a starting point ($\mathscr{A}$) to the target point at $\mathscr{B}$ (Movie S4). Detection region of the swimmers is significantly stifled, such that reduction in ADR exceeds $50\%$ during the trip. This is equivalent to minimal disturbances toward the ambient fluid with $50.4\%$ reduction in MDF. Total time taken for the swimmers to reach the target point is $210 T_s$. The normalized swimming cost is $\bar{C}(x,y) =1 $ for $x/l <50$, and $\bar{C}(x,y) = 2$ for $x/l \geq50$. The optimal (i.e. fastest) pathway from $\mathscr{A}$ to $\mathscr{B}$ is computed through fast marching level-set method, and is shown by a black dashed line. Isolines correspond to $\bar{\mathcal{T}}(x,y)$, that is the minimum time required to reach any point $(x,y)$, starting from point $\mathscr{A}$. Trajectories of the swimmers (tracking the optimal pathway) are also shown by blue, red and green solid lines. The inset represents a specific moment from the trip (as marked by a dashed circle), where schematics demonstrate arrangement of the swimmers at this moment.}
	}
	\label{figS3}
\end{figure}

\hl{Here, we show that through altruistic collaborations, micro-swimmers can also remain stealth while traveling toward a target point or tracking a desired trajectory in such \textit{non-uniform} environments. In particular, we are interested to find an \textit{optimally} concealed traveling swarm, that is the fastest among those with highest possible concealing efficiency -- see e.g. the one presented in Fig. \ref{figM4}(b) passing through a uniform environment. Note that a straight line connecting two points in non-uniform environments no longer represents the fastest pathway between them. Therefore, one needs to first find the optimal (i.e. fastest) pathway from the starting point to the target point. Then, a similar procedure (as outlined in section III-C) can be used to track the specified trajectory. However, there is a caveat here: $\bar{\text{RMS}}_d$ in the objective function ($\mathcal{Z}$), now stands for the normalized root mean square of swimmers' distances from the optimal pathway.}

\hl{As a benchmark, let us consider a simple example of two side-by-side regions (Fig. \ref{figS3}), each with a distinct swimming cost for swimmers. This, for instance, can represent the interface between two distinct liquids. A concealed swarm of three micro-swimmers tracking a prescribed optimal pathway (from $\mathscr{A}$ to $\mathscr{B}$) through such an inhomogeneous environment is represented in Fig. \ref{figS3} (see also Movie S4). At the cost of only 30\% increase in the travel time, compared to the fastest possible swarm, detection region of the swarm is significantly stifled, so that reduction in ADR exceeds $50\%$ during the trip. This is also equivalent to minimally disturbing the ambient fluid with $50.4\%$ reduction in MDF.}

\hl{It also worths noting that by solving the normalized Eikonal equation, i.e. 
$|\nabla \bar{\mathcal{T}}| = \bar{{C}}(\bar{x},\bar{y})$,
using a fast marching level-set method \cite{sethian1996fast, sethian1999level}, one can find $\bar{\mathcal{T}}(\bar{x},\bar{y})$, which is the minimum cost (i.e., the least required time) of reaching to any arbitrary point $(\bar{x},\bar{y})$ in space. Here, the bar signs denote dimensionless quantities and $\bar{C}(\bar{x},\bar{y})$ is the swimming cost at $(\bar{x},\bar{y})$ normalized by $U_s^{-1}$. Values of $\mathcal{T}$ are also normalized by the time scale $T_s = L_s/U_s$. Tracing back from point $\mathscr{B}$ to $\mathscr{A}$, while always moving normal to the isolines of $\bar{\mathcal{T}}$ (see Fig. \ref{figS3}), will then provide the optimal pathway from $\mathscr{A}$ to $\mathscr{B}$ \cite{sethian1999level}. }

\hl{ \section{Expansion of a concealed swarm} }

\begin{figure}[!htb]
	\centering 
	\includegraphics[width=0.49\textwidth]{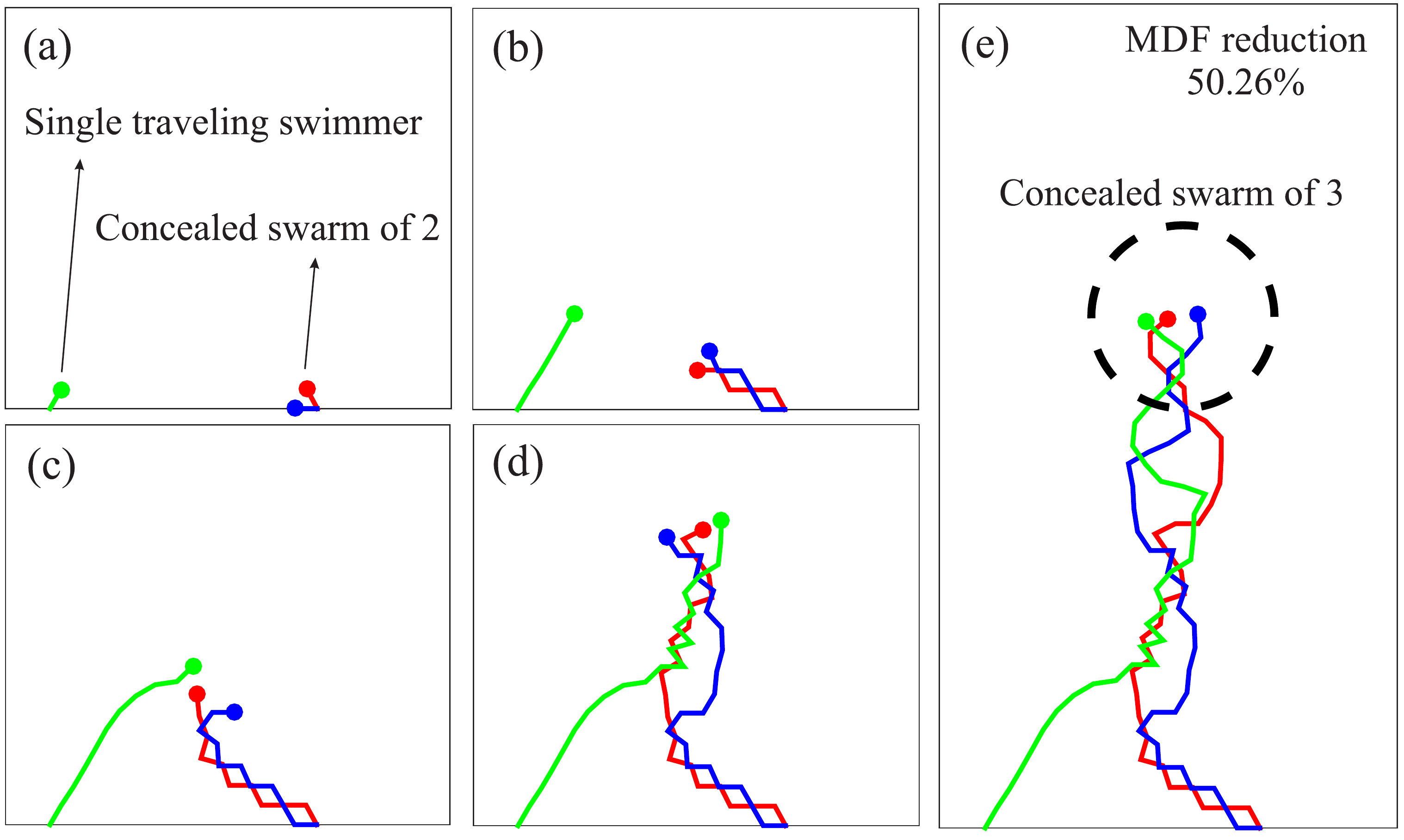}
	\caption {
		\hl{Snapshots of a single traveling swimmer joining a concealed swarm to minimize the overall disturbing flows. (a)-(b) There is a mutual desire to bring the single swimmer into the swarm. (c)-(d) Members trying to form a new optimal swarm arrangement to include the new member in the group. (e) A concealed swarm of three swimmers is formed with more than $50\%$ reduction in MDF.}
	}
	\label{figS2}
\end{figure}

\hl{It is desired for individual swimmers to form a group and collaborate to cancel out each others disturbing effects to the surrounding fluid. Our results further reveal that a traveling concealed swarm can attract nearby individual swimmers (those swimming in its vicinity), ans subsequently expand and re-form into a new larger swarm. To provide further insight, the minimal example is demonstrated in Fig. \ref{figS2} via successive snapshots (a)-(e). It is shown how a single traveling swimmer joins a nearby concealed swarm of two swimmers, and together, they form a new concealed swarm of three swimmers. Note that the only imposed constraint on the motion of swimmers is the upward swimming (c.f. gravitaxis). Relaxing this constraint will simply result in a quasi-random walk of the swarm with no preferred direction.}

%

\hl{ \section{Description of Supplementary Movies} }

\paragraph*{Movie S1}
\hl{The time Evolution of an active concealed swarm of ten swimmers. The agents are initially positioned and oriented randomly (at $t/T_s = 0$). Thereafter, each agent represents a version of run-and-tumble dynamics (with $\tau_r/T_s = 5$), so that to keep the swarm's arrangement within the optimal region of configurations at every instant of time. The immediate value of MDF (computed over a surrounding ring of radius $R/L_s = 100$) is also noted for each of the presented snapshots. Here, the instantaneous position of the swimmers are denoted by colored dots, and gray-scale lines represent their trajectories over time.}

\paragraph*{Movie S2}
\hl{A traveling flock of three micro-swimmers starting from point $\mathscr{A}$ with the intention to reach a target point at $\mathscr{B}$. This traveling swarm is controlled by $\epsilon = 1$ and, as a result, has the highest possible concealing efficiency (i.e. MDF = $49.7\%$). However, the swarm control has no trace of constraints on preferred direction, and thus it never reaches the target. Here, trajectories of the swimmers are shown by blue, green, and red solid lines. The dashed circle mark instantaneous position of the swarm.}

\paragraph*{Movie S3}
\hl{A traveling flock of three micro-swimmers starting from point $\mathscr{A}$ with the intention to reach a target point at $\mathscr{B}$. This traveling swarm is controlled by $\epsilon = 0.5$, and reaches the target in stealth mode (i.e. with the most possible concealing efficiency of MDF = $49.7\%$) in the cost of only 23\% increase in its travel time. Here, trajectories of the swimmers are shown by blue, green, and red solid lines. The dashed circle mark instantaneous position of the swarm.}

\paragraph*{Movie S4}
\hl{A concealed swarm of three micro-swimmers tracking the optimal trajectory in a non-uniform environment from a starting point ($\mathscr{A}$) to the target point at $\mathscr{B}$. Detection region of the swimmers is significantly stifled, such that reduction in ADR exceeds $50\%$ during the trip. This is equivalent to minimally disturbing the ambient fluid with $50.4\%$ reduction in MDF. The total time taken for the swimmers to reach the target point is $210 T_s$. The normalized swimming cost is $\bar{C}(x,y) =1 $ for $x/l <50$, and $\bar{C}(x,y) = 2$ for $x/l \geq50$, where $l$ is the characteristic length of the swimmers (i.e. $L_s$). The optimal (i.e. fastest) pathway from $\mathscr{A}$ to $\mathscr{B}$ is computed through fast marching level-set method, and shown by a black dashed line. Isolines correspond to $\bar{\mathcal{T}}(x,y)$, that is the minimum time required to reach any point $(x,y)$, starting from point $\mathscr{A}$. Trajectories of the swimmers (tracking the optimal pathway) are also shown by blue, red and green solid lines.}


%

\end{document}